# Guided Dispersion Characteristics of Subwavelength Nanoscale Plasmonic Coaxial Lines

Ki Young Kim

*Department of Physics, National Cheng Kung University,*

*1 University Road, Tainan 70101, Taiwan, Republic of China*

The guided dispersion characteristics of subwavelength nanoscale plasmonic coaxial lines are analyzed in detail over a wide optical frequency range below the plasma frequency with a varying gap between the inner and outer plasmonic conductors, providing important preliminary information for understanding the guided mode characteristics of subwavelength nanoscale coaxial lines with plasmonic metals. In particular, when the inner or outer radius is varied, guided modes with similar and dissimilar dispersive features are obtained and discussed. A brief comparison of the propagating modes for the present subwavelength plasmonic coaxial lines and those for a subwavelength plasmonic hole and wire is also made.

**Keywords:** Subwavelength Nanoscale Plasmonic Coaxial Lines, Surface Plasmon Polaritons, Guided Mode Dispersions

## 1. INTRODUCTION

The widespread use of coaxial lines (CXLs), also called coaxial cables and coaxial waveguides, for transmitting radio frequencies (RF) is due to the existence of a fundamental transverse electromagnetic (TEM) mode with no cutoff frequency,[1] which allows the cross-section dimensions to be reduced to a subwavelength diametric scale for efficient RF signal transmission. As such, nanoscale CXLs have also been receiving significant attention for optical frequencies.[2-15] However, in the case of optical frequencies, the permittivity or dielectric constant of metals has a frequency-dependent finite negative value, thereby distinguishing the optical behavior of metals from their behavior with lower frequencies, such as RFs and microwave frequencies, where metals have usually been treated as the perfect electric conductor (PEC) with a negative infinite permittivity. The optical behavior of metals is caused by the interaction between the externally excited photons and the electrons inside the metal, known as the plasmon effect.[16] Due to the finite negative value of the permittivity, the optical waves along the metal surface are tightly confined between the metal surface and the dielectric medium,[17] and referred to as surface plasmon polaritons (SPPs). While a nanoscale CXL, composed of a concentric circular metallic rod in a larger insulating hole with a circular metallic boundary, can support SPPs, the guided characteristics of the electromagnetic waves in a nanoscale CXL are quite distinct from those in a conventional CXL due to the negative dielectric constant of the metal. Nonetheless, despite such differences in the metal behavior for optical and lower frequencies, the properties of the propagating modes of CXLs remain essentially the same, including the existence of a mode with no cutoff frequency, which is why subwavelength nanoscale plasmonic CXLs are still considered an important guiding structure for optical frequencies.

Studies on optical coaxial structures first began in the area of near-field scanning optical microscopes (NSOMs), where subwavelength coaxial tips were investigated to enhance the NSOM resolution.[2-5] More recently, enhanced optical transmission (EOT) through a subwavelength coaxial aperture has also become an area of intensive research,[5-9] plus studies on the guided modes of optical CXLs[7,10-15] have produced good optical field concentrations in subwavelength CXL structures. However, despite such extensive research, the guided dispersion characteristics of subwavelength nanoscale CXLs with plasmonic conductors still need to be investigated further. In particular, understanding the effect of varying the gap on the guided dispersion over a wide frequency regime is still insufficient, despite its significance as a key design parameter of a subwavelength nanoscale plasmonic CXL and key component in understanding the propagation of subwavelength light in this waveguide.

Accordingly, this paper theoretically examines the propagation characteristics of subwavelength nanoscale plasmonic CXLs in terms of their guided dispersion characteristics over a wide frequency range below the plasma frequency. As a result, an overall and revealing picture of the guided modes is obtained and discussed with an emphasis on the effect of varying the gap width between the inner and outer plasmonic conductors on the dispersion characteristics. It is hoped that the present findings will provide new insights into subwavelength nanoscale plasmonic coaxial transmission lines and their secondary applications, such as superfocusing, NSOMs, and EOT.

## 2. PLASMONIC COAXIAL LINE STRUCTURE AND ITS CHARACTERISTIC EQUATIONS

Figure 1 shows a cross-sectional schematic view of the plasmonic CXL and cylindrical coordinates considered in this paper. Regions 1 and 3 are the plasmonic conductors, region 2 is the air region, and $a$ and $b$ are the radii of the inner and outer conductors, respectively. The light propagation is in the $+z$-direction, *i.e.* perpendicular to the cross section presented in Figure 1. The relative permeabilities and dielectric constant in the air region are assumed to be unity, *i.e.* $\mu_{r1} = \mu_{r2} = \mu_{r3} = \varepsilon_{r2} = 1.0$. For the metal regions, silver was used, and its dielectric constant can be expressed using the simple Drude model of $\varepsilon_{r1} = \varepsilon_{r3} = 1 - \omega_p^2 / [\omega(\omega - j\gamma)]$, where $\omega(= 2\pi f)$ is the angular operating frequency ($f$ is the operating frequency), and $\omega_p$ and $\gamma$ are the angular plasma frequency and collision frequency, respectively. The plasma frequency of silver is $f_p = \omega_p / 2\pi = 2175 \text{THz}$,[18] and the collision frequency has been disregarded for simplicity. As such, the dispersion of the dielectric constant of silver is shown in Figure 2. Below the plasma frequency, the dispersive dielectric constant of silver becomes negative, while above the plasma frequency it becomes positive, which allows the positive refractive index of silver to be determined as shown in Figure 2, although this is not the main concern here. Figure 2 also includes another critical frequency, called the plasma resonance frequency, *i.e.* $f_r = f_p / \sqrt{2} = 1537.96 \text{THz}$, where the absolute value of the dielectric constant of silver becomes the same as that of the free space, which plays an important role in the guided dispersion characteristics of subwavelength nanoscale plasmonic CXLs, as shown in the next section. Using this simple Drude model of silver, the dielectric constant remains valid in near- and far-infrared regions. For a more accurate description of the optical behavior in higher frequency regimes, such as visible and ultraviolet, additional Lorentzian terms are needed,[19] yet this can obscure the more important effect of the dispersive negative dielectric constant. Besides, recent studies have also shown important results for a similar structure and dimensions with a wide frequency regime using this simple Drude model for silver.[20-22] Thus, it was decided to use the simple Drude model with a wide optical frequency regime while noting the abovementioned limitations. As a result, a clear picture of the guided dispersion of subwavelength nanoscale plasmonic CXLs was obtained, which will be discussed in detail in the next section.

As the field distributions of the SPP mode in this waveguide are all evanescent in the transverse direction ($+r$), and only the transverse magnetic (TM) mode can propagate in the form of the SPP mode, the electric fields of the propagation direction in each region can be expressed as follows:

$$E_{z1} = AI_m(k_1 r)\cos m\theta \exp\left[j(\omega t - \beta z)\right] \quad (r<a) \quad \text{in region 1} \quad (1)$$

$$E_{z2} = \left[BI_m(k_2 r) + CK_m(k_2 r)\right]\cos m\theta \exp\left[j(\omega t - \beta z)\right] \quad (a<r<b) \quad \text{in region 2} \quad (2)$$

$$E_{z3} = DK_m(k_3 r)\cos m\theta \exp\left[j(\omega t - \beta z)\right] \quad (r>b) \quad \text{in region 3} \quad (3)$$

where $A$, $B$, $C$, and $D$ are the constants associated with the amplitudes between the field quantities in each region, $I_m(\cdot)$ and $K_m(\cdot)$ are $m$th order modified Bessel functions of the first and second kind, respectively, where $m$ is the azimuthal eigenvalue, and $\beta$ and $k_i (i=1,2,3)$ are the propagation constants in the propagation and transverse directions, respectively, and correlated as $k_i = \left(\beta^2 - k_0^2 \mu_{ri}\varepsilon_{ri}\right)^{1/2}$, where $k_0$ is the propagation constant of the free space. The evanescent field in region 2 imposed the TM wave on the slow wave, *i.e.* $\beta > k_0 \left(\mu_{r2}\varepsilon_{r2}\right)^{1/2}$, which is one of the important properties of the SPP mode.

If concern is restricted to the azimuthally invariant mode, *i.e.* $m=0$, the azimuthal magnetic field components in each region, meaning the field components tangential to the boundaries at $r=a$ and $r=b$, can be obtained from the above field components in the propagating direction using Equations (1) to (3) as follows:

$$H_{\theta 1} = A \frac{j\omega\varepsilon_0\varepsilon_{r1}}{k_1} I_0'(k_1 r)\exp\left[j(\omega t - \beta z)\right] \quad (r<a) \quad \text{in region 1} \quad (4)$$

$$H_{\theta 2} = \frac{j\omega\varepsilon_0\varepsilon_{r2}}{k_2}\left[BI_0'(k_2 r) + CK_0'(k_2 r)\right]\exp\left[j(\omega t - \beta z)\right] \quad (a<r<b) \quad \text{in region 2} \quad (5)$$

$$H_{\theta 3} = D\frac{j\omega\varepsilon_0\varepsilon_{r3}}{k_3} K_0'(k_3 r)\exp\left[j(\omega t - \beta z)\right] \quad (r>b) \quad \text{in region 3} \quad (6)$$

where prime denotes the differentiation with respect to $r$, and $\varepsilon_0$ is the permittivity of the free space. When following the standard steps of the boundary value problem using Equations (1) to (6) at $r = a$ and $r = b$ in the cylindrical coordinate system, the characteristic equation for the SPP mode of the plasmonic CXL can be derived as follows:

$$P_{SPP}S_{SPP} - Q_{SPP}R_{SPP} = 0, \quad (7)$$

where $P_{SPP} = I_0(k_2 a)(\varepsilon_{r2}\xi_2 - \varepsilon_{r1}\xi_1)$ $\quad S_{SPP} = K_0(k_2 b)(\varepsilon_{r2}\xi_6 - \varepsilon_{r3}\xi_4)$

$Q_{SPP} = K_0(k_2 a)(\varepsilon_{r2}\xi_3 - \varepsilon_{r1}\xi_1)$ $\quad R_{SPP} = I_0(k_2 b)(\varepsilon_{r2}\xi_5 - \varepsilon_{r3}\xi_4)$

with $\xi_{SPP}^1 = \frac{1}{k_1 a}\frac{I_0(k_1 a)}{I_0'(k_1 a)}$ $\quad \xi_{SPP}^2 = \frac{1}{k_2 a}\frac{I_0'(k_2 a)}{I_0(k_2 a)}$ $\quad \xi_{SPP}^3 = \frac{1}{k_2 a}\frac{K_0'(k_2 a)}{K_0(k_2 a)}$

$\xi_{SPP}^4 = \frac{1}{k_3 b}\frac{K_0'(k_3 b)}{K_0(k_3 b)}$ $\quad \xi_{SPP}^5 = \frac{1}{k_2 b}\frac{I_0'(k_2 b)}{I_0(k_2 b)}$ $\quad \xi_{SPP}^6 = \frac{1}{k_2 b}\frac{K_0'(k_2 b)}{K_0(k_2 b)}.$

Meanwhile, there is also another way of propagation in the plasmonic CXL, in which the field distribution in region 2 is oscillating rather than evanescent. This is similar to the guidance of a higher order mode in conventional CXLs in a fast wave region, *i.e.* $\beta < k_0(\mu_{r2}\varepsilon_{r2})^{1/2}$, referred to as the CXL mode in this paper. In this case, the electric field component in the propagating direction in region 2 can be rewritten as $E_{z2} = \left[BJ_0(k_2 r) + CY_0(k_2 r)\right]\exp\left[j(\omega t - \beta z)\right]$, where $J_0(\cdot)$ and $Y_0(\cdot)$ are 0-th order Bessel functions of the first and second kind, respectively. Note that the field expressions in the other regions remain the same as for the previous SPP mode due to the negative character of the dielectric constant in the plasmonic conductor regions and the radiation condition. As such, the resultant characteristic equation for the CXL mode can also be obtained using the aforementioned procedure for the SPP mode as follows:

$$P_{CXL} S_{CXL} - Q_{CXL} R_{CXL} = 0, \tag{8}$$

where $P_{CXL} = J_0(k_2 a)(\varepsilon_{r2}\xi_2 + \varepsilon_{r1}\xi_1)$ $\quad S_{CXL} = Y_0(k_2 b)(\varepsilon_{r2}\xi_6 + \varepsilon_{r3}\xi_4)$

$Q_{CXL} = Y_0(k_2 a)(\varepsilon_{r2}\xi_3 + \varepsilon_{r1}\xi_1)$ $\quad R_{CXL} = J_0(k_2 b)(\varepsilon_{r2}\xi_5 + \varepsilon_{r3}\xi_4)$

with $\xi_{CXL}^1 = \dfrac{1}{k_1 a}\dfrac{I_0(k_1 a)}{I_0'(k_1 a)}$ $\quad \xi_{CXL}^2 = \dfrac{1}{k_2 a}\dfrac{J_0'(k_2 a)}{J_0(k_2 a)}$ $\quad \xi_{CXL}^3 = \dfrac{1}{k_2 a}\dfrac{Y_0'(k_2 a)}{Y_0(k_2 a)}$

$\xi_{CXL}^4 = \dfrac{1}{k_3 b}\dfrac{K_0'(k_3 b)}{K_0(k_3 b)}$ $\quad \xi_{CXL}^5 = \dfrac{1}{k_2 b}\dfrac{J_0'(k_2 b)}{J_0(k_2 b)}$ $\quad \xi_{CXL}^6 = \dfrac{1}{k_2 b}\dfrac{Y_0'(k_2 b)}{Y_0(k_2 b)}.$

The characteristic equations of Equations (7) and (8) represent the guided dispersion characteristics of plasmonic CXLs. The solutions of Equation (8) are continuously connected to those of Equation (7) at the border of slow and fast waves ($\beta = k_0$), which will be demonstrated shortly.

## 3. NUMERICAL RESULTS AND DISCUSSION

Figure 3 shows the normalized propagation constant ($\beta/k_0$) versus the operating frequency (*f*) when varying the outer conductor radius with a fixed inner conductor radius of $a = 50\text{nm}$, thereby demonstrating the guided dispersion characteristics of subwavelength nanoscale plasmonic CXLs. The horizontal dotted and vertical dot and dashed lines represent the border of the slow ($\beta > k_0$) and fast ($\beta < k_0$) waves, i.e. $\beta = k_0$ or $\beta/k_0 = 1.0$, and the plasma resonance frequency ($f_r = 1537.96\text{THz}$), respectively. The solution of Equation (7) for the SPP modes generated two distinctive modes in the slow wave region, represented by the $TM_{00}$ and $TM_{01}$ modes shown in Figure 3(a). The guided dispersion characteristics of the mode with both slow and fast waves were similar to those of the $TM_{01}$ mode of a subwavelength plasmonic hole,[23] indicating that the guided dispersion characteristics of this mode mainly resulted from the outer boundary of the plasmonic conductor, as a fast CXL mode cannot exist in a open structure, hence it was named the $TM_{01}$ mode. Meanwhile, the other mode, which existed below the plasma resonance frequency and did not have a low-frequency cutoff, is not observed in a subwavelength plasmonic hole,[23] apparently due to the introduction of an additional inner concentric plasmonic conductor in the hole with a plasmonic conductor boundary. While the propagation constant of the TEM mode for conventional air-filled PEC CXLs is the same as the speed of light,[1] *i.e.* $\beta = k_0$ [horizontal dotted line], the propagation constant of this mode showed slow wave characteristics at any operating frequency, due to the finite value of the negative dielectric constant of the plasmonic conductor, and approached the $\beta = k_0$ line as the operating frequency decreased. Thus, due to its resemblance at a lower frequency regime to the TEM mode of a conventional CXL in the RF band, this mode was named the $TM_{00}$ mode. As this TEM-like $TM_{00}$ mode does not suffer from a low-frequency cutoff, which means that the guided mode can propagate regardless of a small gap between the inner and outer plasmonic conductors, it can be used for subwavelength guidance for optical frequencies, similar to the intensive use of conventional CXLs for the RF band, as previously mentioned. This kind of quasi-TEM mode can also be found in other plasmonic waveguiding systems with multiple plasmonic conductors, such as a plasmonic microstrip line,[24] plasmonic coplanar line,[25] and plasmonic parallel plate waveguide (PPW),[26] and its existence can be explained using classical electromagnetic field theory.[27]

In Figure 3(a), as the outer conductor radius decreased from $b = 150$ to 60nm, the cutoff frequency, *i.e.* the frequency at $\beta/k_0 = 0$, of the $TM_{01}$ mode shifted toward a higher frequency regime and resonated at $f_r = 1537.96 \text{THz}$. Thus, the $TM_{01}$ modes with wider gaps, such as $b = 150$, 120, and 100nm, had forward waves with positive slopes on the dispersion curves, while those with narrower gaps, such as $b = 60$ and 80nm, had backward waves with negative slopes, representing a negative group velocity.[28] This also coincided with the case of a subwavelength plasmonic hole without an inner plasmonic conductor,[23] confirming that the guided dispersion characteristics of the $TM_{01}$ mode of the subwavelength plasmonic CXL were mainly due to the effects of the outer plasmonic conductor. The $TM_{01}$ mode for $b = 100$nm only existed within a narrower frequency range when compared to the other cases. This was due to the specific choice of $a = 50$nm and $b = 100$nm for the geometrical dimensions, showing that the resonance of the guided dispersion characteristics is not only related to the plasma frequency of the plasmonic conductor but also to the geometrical parameters. When a slightly smaller value than $b = 100$nm was used as the outer radius, the guided dispersion curve became almost vertical like the vertical dot and dashed line for the plasmonic resonance, which may have potential application for highly sensitive optical sensors with a variety of phase velocities ($k_0/\beta$) from fast to slow waves when a proper loss compensation scheme can be implemented.[29]

Meanwhile, the normalized propagation constant of the $TM_{00}$ mode maintained forward waves until it reached a very high value at the plasma resonance frequency, as shown in Figure 3(b). Plus, it also increased as the radius of the outer conductor decreased from $b = 150$ to 60nm, because more optical power was confined at the interface as the gap between the inner and outer conductor decreased. It is interesting to note that the forward $TM_{01}$ modes of $b = 100$, 120, and 150nm had a higher normalized propagation constant as the outer radius increased, while the $TM_{00}$ modes had lower normalized propagation constants. As such, in the extreme case of an infinite outer radius ($b = \infty$), corresponding to a bare plasmonic wire with $a = 50$nm, the two dispersion curves could be expected to combine into a single mode for the plasmonic wire, as shown in Figure 3(b), which is also a slow wave with no cutoff frequency.[30] This slow wave of the plasmonic wire could also be regarded as a $TM_{00}$ mode, since the effect of the outer conductor boundary, which is the main factor affecting the existence of the $TM_{01}$ mode, is absent, plus it followed the tendency of the $TM_{00}$ modes of the subwavelength plasmonic CXLs.

Figure 4 shows the guided dispersion characteristics of the subwavelength nanoscale plasmonic CXL when varying the inner conductor radius with the outer radius of the conductor fixed at $b = 100$nm, which was the reverse of the case in Figure 3. For the $TM_{01}$ mode, as the radius of the inner conductor increased, the low-frequency cutoff shifted toward a higher frequency regime, and the modes also resonated at the plasma resonance frequency of $f_r = 1537.96$THz. Thus, the $TM_{01}$ modes with a wider gap had forward waves, while those with a relatively large inner conductor, such as $a = 70$ and 90nm, had backward waves, which was consistent with the previous cases shown in Figure 3(a). The $TM_{01}$ mode of a subwavelength plasmonic hole[23] with $b = 100$nm, *i.e.* $a = 0$nm for the CXL, is also shown in Figure 4(a) for the purpose of comparison. There was no $TM_{00}$ mode for the subwavelength plasmonic hole due to the absence of an inner conductor. The normalized propagation constant and cutoff frequency for the $TM_{01}$ mode for the $a = 0$nm case were slightly higher and lower than those for the $a = 10$nm case, respectively, and showed a well-behaved tendency with the other dimensions of the inner conductor radius, which could be taken as evidence that the mode classification of the subwavelength CXL in this study was proper. Thus, it can be concluded that the existence of the $TM_{00}$ and $TM_{01}$ modes was mainly dependent on the presence of the inner and outer plasmonic conductors, respectively, based on the results discussed for Figures 3(b) and 4(a), and that the guided modes of the subwavelength plasmonic CXL can be regarded as a combination of the guided modes in the case of a bare plasmonic wire and plasmonic hole.

Figure 4(b) shows the guided dispersion characteristics of the subwavelength nanoscale plasmonic CXL for the $TM_{00}$ mode, which look quite different from the case in Figure 3(b). In Figure 4(b), the $TM_{00}$ mode with a smaller radius had a higher normalized propagation constant, *i.e.* the normalized propagation constant for the case of $a = 10$nm was higher than those for the other guided modes. However, the normalized propagation constant of $a = 90$nm was also higher than the guided modes for the cases of $a = 30, 50,$ and $70$nm. Although this initially appears contradictory, it can be qualitatively explained using two different field confinement mechanisms: a radius with a higher curvature (small radius) can confine more light energy on the surface,[31] thereby resulting in a higher normalized propagation constant for the case of $a = 10$nm, and the guided energy of the subwavelength light can also be well confined in the case of a small gap, such as $a = 90$nm. The latter case is the same as the case in Figure 3(b). Therefore, the guided dispersion characteristics in the cases of $a = 30, 50,$ and $70$nm, which look rather unusual, resulted from a combination of the two field confinement mechanisms with the dispersive characteristics of silver shown in Figure 2. Although this abnormal guided dispersion is not likely to be found in the case of relatively larger gap dimensions between the two plasmonic conductors or a narrow operating frequency band, it could affect the propagation or concentration performance of subwavelength nanoscale plasmonic CXLs utilized in subwavelength focusing applications using tapered structures.

## 4. CONCLUSION

This paper theoretically investigates the guided dispersion characteristics of a subwavelength nanoscale CXL over a wide frequency range below and above the plasma resonance frequency of the metal and over slow and fast wave regions. When the inner or outer conductor radii were varied, similar and dissimilar guided dispersion characteristics were obtained and discussed. The existence of the $TM_{00}$ and $TM_{01}$ modes was considered to be mainly influenced by the presence of the inner and outer plasmonic conductor boundaries, respectively, and the guided modes of the present structure were a combination of the guided modes of a bare plasmonic wire and plasmonic hole without an inner conductor. In addition, unusual guided dispersion characteristics occurred with the $TM_{00}$ mode when the inner conductor radius was varied with a fixed outer radius, and there characteristics were explained using two different field or energy confinement mechanisms combined with the frequency dependent property of the metal. While air was used in the present study as the filling material between the two plasmonic conductors, choosing a different dielectric material would produce shifts in the dispersion curves, yet the general findings presented in this paper would remain valid.

Generally, plasmonic waveguides are known to be highly lossy, however, this was not considered important in the present study, as it is not a serious problem when plasmonic waveguides are used as guiding media for very short distances and/or as part of highly integrated nanoscale devices. Moreover, incorporating optically active media[32, 33] as a filling material between two plasmonic conductors in a subwavelength nanoscale plasmonic CXL could be a solution for mitigating this problem. Although a gain-compensated low-loss subwavelength propagation mechanism using electrical pumping for a gap-filled optically active medium looks quite challenging at present, this idea could be helpful in developing future applications of subwavelength nanoscale plasmonic CXLs.

Finally, the guided dispersion characteristics of a plasmonic PPW using different plasmonic metals with different bulk plasma frequencies in an optical frequency regime have already been investigated by the current author[34]. Thus, from the present results and results in Ref. 34, it is expected that the use of different plasmonic metals in subwavelength plasmonic CXLs will cause different guided dispersion characteristics due to the combination of different plasmonic dispersions from different metals and their structures, *i.e.* different radii for the curvature of the inner and outer plasmonic conductors. Such results will be reported elsewhere.

**Figure Captions**

**Fig. 1.** Cross-sectional schematic view of subwavelength nanoscale plasmonic coaxial line.

**Fig. 2.** Dielectric constant of silver using simple Drude model.

**Fig. 3.** Guided dispersion characteristics of subwavelength nanoscale plasmonic coaxial line with fixed inner conductor radius ($a$=50nm).

**Fig. 4.** Guided dispersion characteristics of subwavelength nanoscale plasmonic coaxial line with fixed outer conductor radius ($b$=100nm).

**Figure 1**

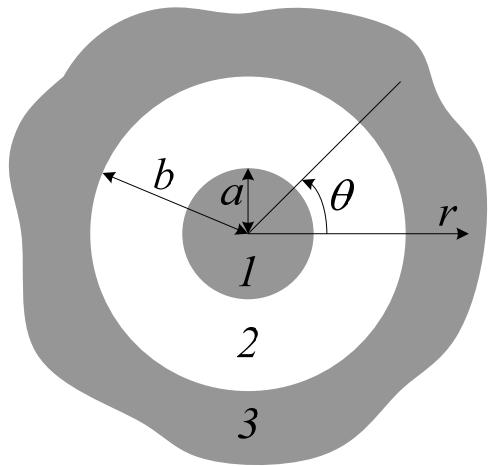

**Figure 2**

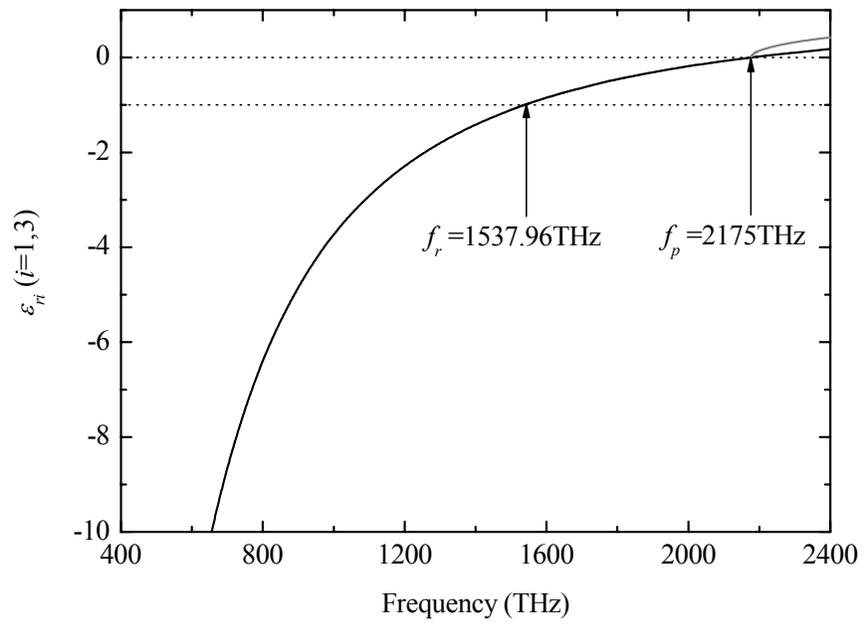

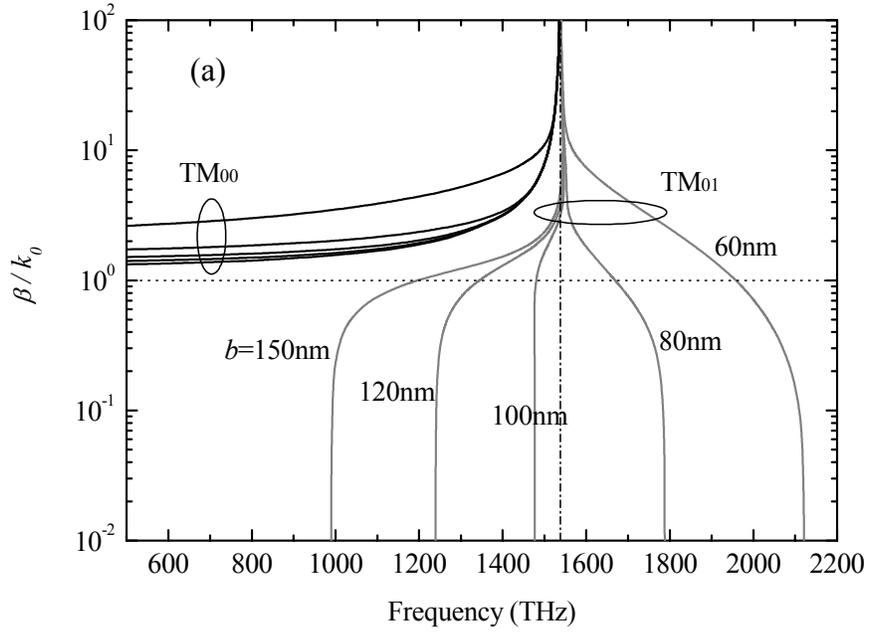

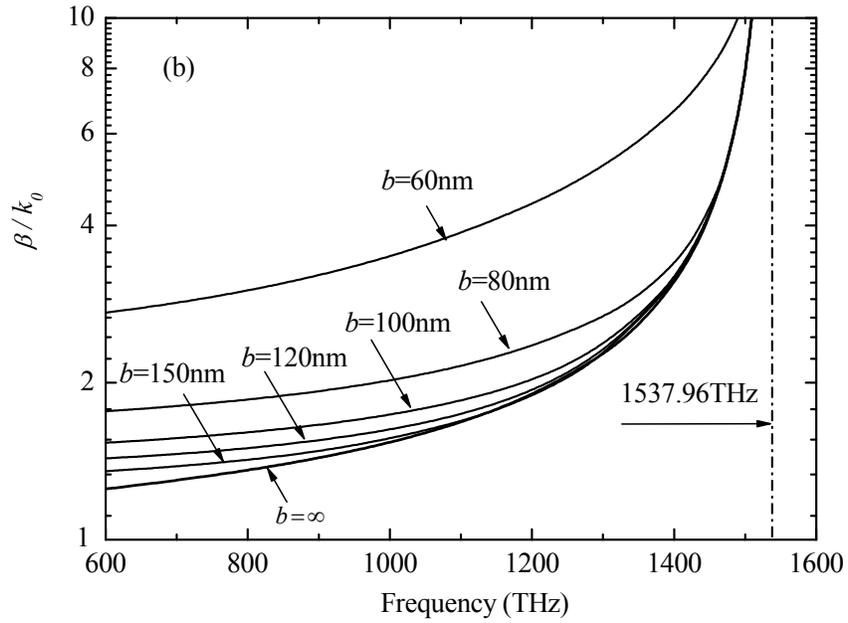

**Figure 4**

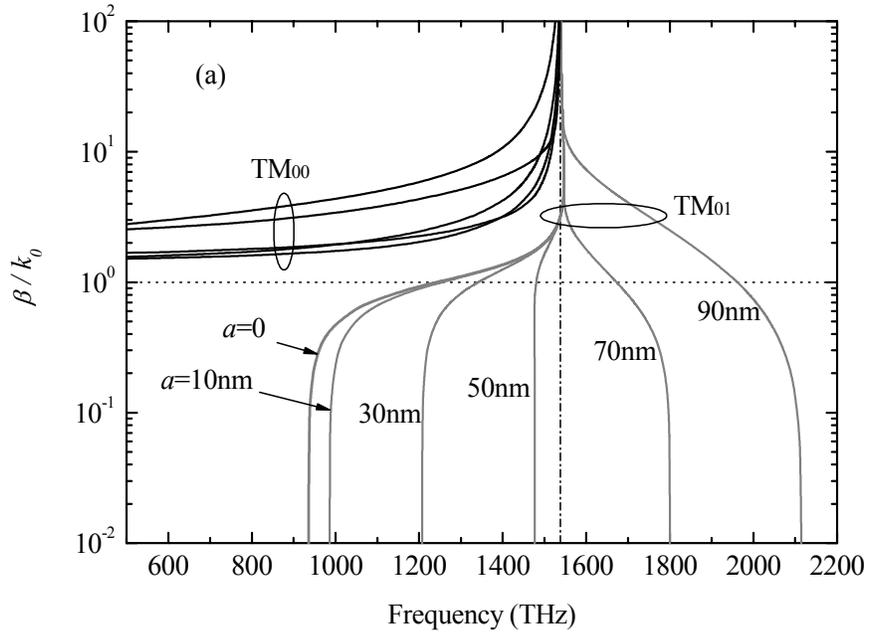

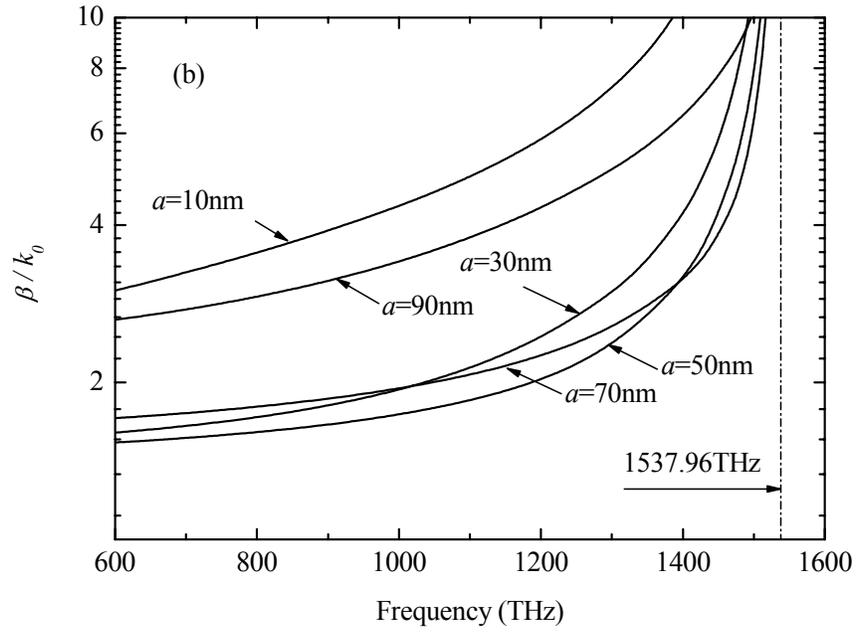